\documentclass{aa}
\usepackage{graphicx}
\usepackage{txfonts}
\begin{document}
\title{
	A possible age-metallicity relation in the 
	Galactic thick disk?\thanks{Based on data obtained with
	the Hipparcos satellite}
	}

\author{
	T. Bensby \and
	S. Feltzing \and	
	I. Lundstr\"om
	}

\offprints{Thomas Bensby}

\institute{
	Lund Observatory, 
	Box 43, 
	S-221 00 Lund\\
        \email{thomas@astro.lu.se, sofia@astro.lu.se, ingemar@astro.lu.se}		}

\date{
	Received 30 December 2003; Accepted 21 March 2004
	}

\abstract{
	A sample of 229 nearby thick disk stars has been
	used to investigate the existence of an age-metallicity relation (AMR)
	in the Galactic thick disk. The results indicate that that there is
	indeed an age-metallicity relation present in the thick disk. 
	By dividing the stellar sample into sub-groups,
	separated by 0.1\,dex in metallicity, we show that the median age
	decreases by about 5--7\,Gyr when going
	from [Fe/H]\,$\approx -0.8$ to [Fe/H]\,$\approx -0.1$. Combining
	our results with our newly published $\alpha$-element trends
	for a local sample of thick disk stars, that show
	signatures from supernovae type Ia (SN\,Ia), 
	we can here draw the conclusion
	that the time-scale for the peak of the SN\,Ia rate is of the order
	3--4 Gyr in the thick disk. The tentative evidence for
	a thick disk AMR that we present here also has implications for 
	the thick disk
	formation scenario; star-formation must have been an ongoing
	process for several billion years. This is further discussed here
	and appear to strengthen the hypothesis that the thick disk 
	originates from a merger event with a companion galaxy that puffed 
	up a pre-existing thin disk.
\keywords{
	(Stars:) Hertzsprung-Russell (HR) and C-M diagrams --
	Stars: kinematics --
	Galaxy: disk --
	Galaxy: kinematics and dynamics --
	(Galaxy:) solar neighbourhood --
	Galaxy: formation 
	}}

\maketitle
%
\section{Introduction}

Chemical evolution of stellar populations is an important part of
any model of galaxy formation and evolution. Many studies in the
past decades show how we are able to further refine and constrain
models of Galactic chemical evolution by combining kinematics and 
elemental abundances of local dwarf stars (e.g. 
Chiappini et al.~\cite{chiappini}; 
Matteucci~\cite{matteucci}; 
Edvardsson et al.~\cite{edvardsson}; 
Feltzing \& Gustafsson~\cite{feltzing1998};
Feltzing et al.~\cite{feltzing2003}; 
Bensby et al.~\cite{bensby}; 
Reddy et al.~\cite{reddy}).
As evidenced by the cited articles our understanding of chemical 
evolution is mainly driven by the studies in the solar neighbourhood 
but have far 
reaching impact for our interpretation of integrated light from 
other galaxies (e.g. Matteucci~\cite{matteucci}).

However, it is not only the elemental abundances and kinematics of 
the stars that are of importance when we want to further improve the models
of galaxy formation and evolution but also the ages of the stars (see e.g. 
Edvardsson et al.~\cite{edvardsson}; Raiteri et al.~\cite{raiteri}; 
Pilyugin \& Edmunds~\cite{pilyugin}; Berczik~\cite{berczik}). 

Many studies have found there to be a clear relation between the ages
and the metallicities of the solar neighbourhood disk stars
(Twarog~\cite{twaroga}, \cite{twarogb}; Rocha-Pinto et al.~\cite{rochapinto}; 
Meusinger et al.~\cite{meusinger}). In contrast to this 
Edvardsson et al.~(\cite{edvardsson}) found no particular evidence for an 
age-metallicity relation in the Galactic disk in the solar neighbourhood and 
Feltzing et al.~(\cite{feltzing}) confirmed this. 
Feltzing et al.~(\cite{feltzing}) also showed how dangerous selection effects 
could be and how an artificial age-metallicity relation can be created 
(see their Figs.~13 and 14).

Gilmore \& Reid~(\cite{gilmore}) showed that our galaxy is host to two 
kinematically distinct disk structures. The ``new'' disk was dubbed the 
thick disk and was found to have a mean metallicity around 
$-0.6$\,dex (Wyse \& Gilmore~\cite{wyse}) and a scale-height of 
800--1300\,pc (e.g. Reyl\'e \& Robin~\cite{reyle}; Chen~\cite{chen2}) 
while the thin disk has a mean metallicity of around
$-0.1$\,dex and a scale height of 100--300 pc 
(e.g. Gilmore \& Reid~\cite{gilmore}; Robin et al.~\cite{robin}). 
Recent studies have shown that stars selected to 
belong to either the thin or the thick disk show different 
trends for the elemental abundances (e.g. Fuhrmann~\cite{fuhrmann};
Feltzing et al.~\cite{feltzing2003}; Bensby et al.~\cite{bensby};
Bensby et al.~\cite{bensby_syre};
Reddy et al.~\cite{reddy}; Prochaska et al.~\cite{prochaska};
Mashonkina \& Gehren~\cite{mashonkina}).

The question then arises: could it be so that the lack of a relation
between ages and metallicities for stars in the solar neighbourhood
is in fact a population effect? That is, are we looking at a mixture
of stars from (at least) two populations that have different 
chemical enrichment histories? 

It thus appears natural to, yet again, revisit the question of an 
age-metallicity relation in the solar neighbourhood. In the study 
presented here we will address the question of a relation between ages and 
metallicities for stars that are kinematically selected to resemble the
thick disk closely.

The paper is organized as follows: in Sect.~\ref{sect:kin} we describe the
stellar sample and the kinematical selection criteria and
investigates if there are potential biases present in the sample.
In Sect.~\ref{sect:alpha} we describe the choice of 
$\alpha$-enhancement used in the isochrones when deriving the stellar ages.
In Sect.~\ref{sec:amr} we derive ages from
stellar isochrone fitting and find that there is a possible
age-metallicity relation present in the thick disk. In 
Sect.~\ref{sec:discussion} we discuss the implications this tentative
age-metallicity relation have on the star-formation history of the thick disk, 
on the time-scale of SN\,Ia rate in the thick disk, and on our understanding 
of the origin and evolution of the thick disk. 
Finally, in Sect.~\ref{sec:summary} we give a short summary.

\section{Stellar sample}
\label{sect:kin}

The stellar sample has been selected on purely kinematical grounds
(see also Bensby et al.~\cite{bensby}; Bensby et al.~submitted). We have 
assumed that the Galactic space velocities ($U_{\rm LSR}$, $V_{\rm LSR}$, and 
$W_{\rm LSR}$) for the stellar populations in the solar neighbourhood 
all can be described by Gaussian 
distributions. For each star (with its specific combination of $U_{\rm LSR}$, 
$V_{\rm LSR}$, and $W_{\rm LSR}$) it is then possible to calculate the 
probabilities that it belongs to either the thin disk ($D$), thick disk ($TD$), 
or the halo ($H$). These can then be used to, for each star, calculate the 
``relative probabilities" $TD/D$ and $TD/H$. When doing this, the fraction 
(normalization) of the 
three components in the solar neighbourhood must be taken into 
account. The final relationship is (see Bensby et al.~\cite{bensby}):
\begin{equation}
        P = X \cdot k \cdot \exp\left(
        -\frac{U^{2}_{\rm LSR}}{2\,\sigma_{\rm U}^{2}}
        -\frac{(V_{\rm LSR} - V_{\rm asym})^{2}}{2\,\sigma_{\rm V}^{2}}
        -\frac{W^{2}_{\rm LSR}}{2\,\sigma_{\rm W}^{2}}\right),
\label{eq:probabilities}
\end{equation}

\noindent
where
\begin{equation}
        k = \frac{1}{(2\pi)^{3/2}\,\sigma_{\rm U}
                                 \,\sigma_{\rm V}
                                 \,\sigma_{\rm W}},
\end{equation}

\noindent
normalizes the expression; $\sigma_{\rm W}$, $\sigma_{\rm V}$,
$\sigma_{\rm W}$ are the characteristic velocity dispersions; $V_{\rm asym}$
is the asymmetric drift; and $X$ is the observed fraction of stars in the
solar neighbourhood for each population. 
The values for the velocity dispersions and the asymmetric drifts are taken
from Bensby et al.~(\cite{bensby}) and the values for the normalizations
in the solar neighbourhood from Bensby et al.~(submitted) (see also discussion 
below). All values are given in Table~\ref{tab:galactic}.

\begin{table}
\centering
\caption{
        Characteristic velocity dispersions
        ($\sigma_{\rm U}$, $\sigma_{\rm V}$, and $\sigma_{\rm W}$) in the thin
        disk, thick disk, and stellar halo, used in
        Eq.~(\ref{eq:probabilities}). $X$ is the observed fraction of stars for
        the populations in the solar neighbourhood and $V_{\rm asym}$ is the
        asymmetric drift (values taken from Bensby et al.~\cite{bensby} and
	Bensby et al.~submitted).
        }
\begin{tabular}{llcccr}
\hline \hline\noalign{\smallskip}
        & $X$
        & $\sigma_{\rm U}$
        & $\sigma_{\rm V}$
        & $\sigma_{\rm W}$
        & $V_{\rm asym}$ \\
        &
        & \multicolumn{4}{c}{---------- [km~s$^{-1}$] ----------}     \\
\noalign{\smallskip}
\hline\noalign{\smallskip}
   Thin disk (D)   & 0.90   & $~~35$  & 20    & 16    & $-15$    \\
   Thick disk (TD) & 0.10   & $~~67$  & 38    & 35    & $-46$    \\
   Halo (H)        & 0.0015 & $160$   & 90    & 90    & $-220$   \\
\hline
\end{tabular}
\label{tab:galactic}
\end{table}
\begin{figure}
\centering
\resizebox{\hsize}{!}{
        \includegraphics[bb=18 144 592 580,clip]{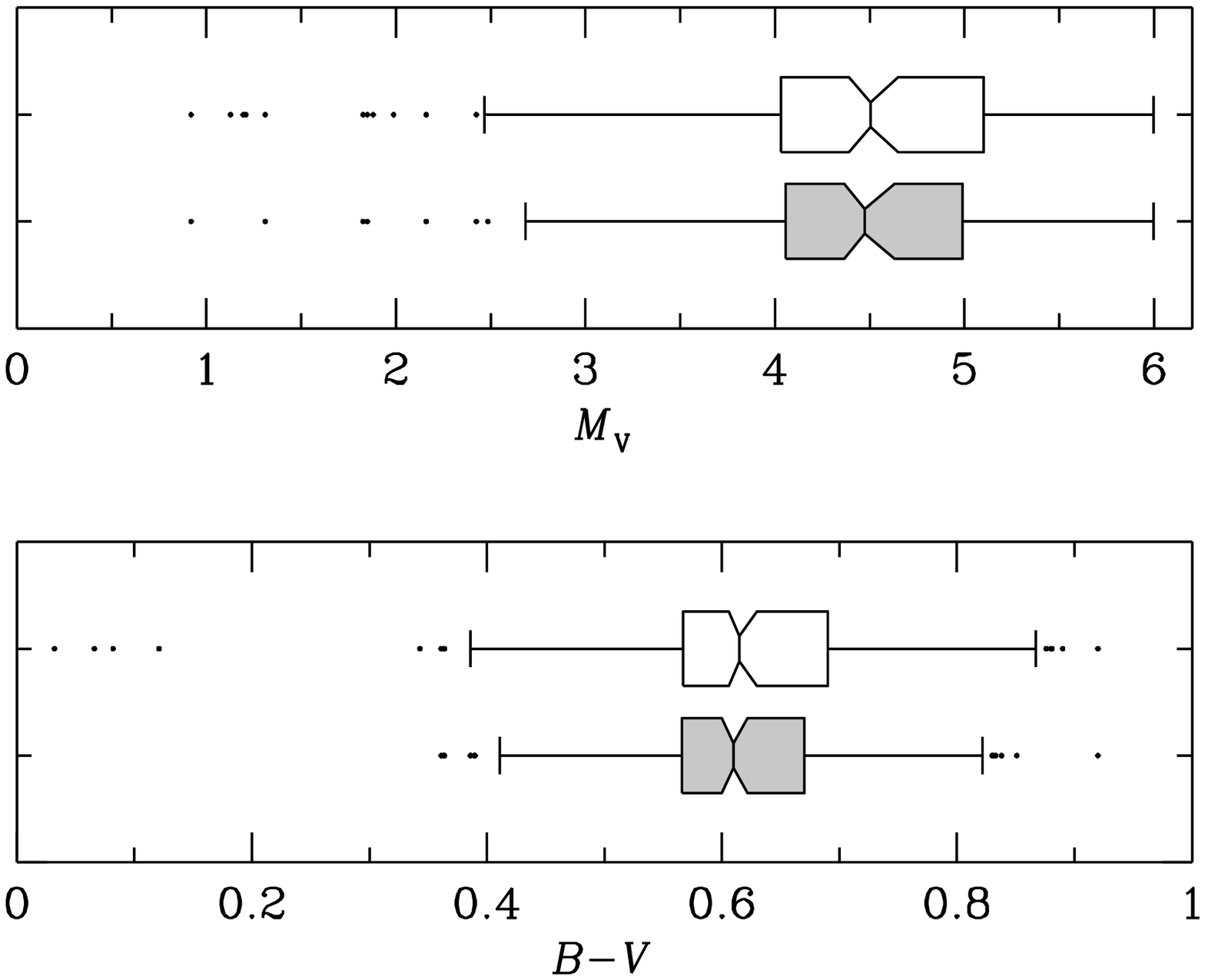}}
\resizebox{\hsize}{!}{
        \includegraphics[bb=18 144 592 607,clip]{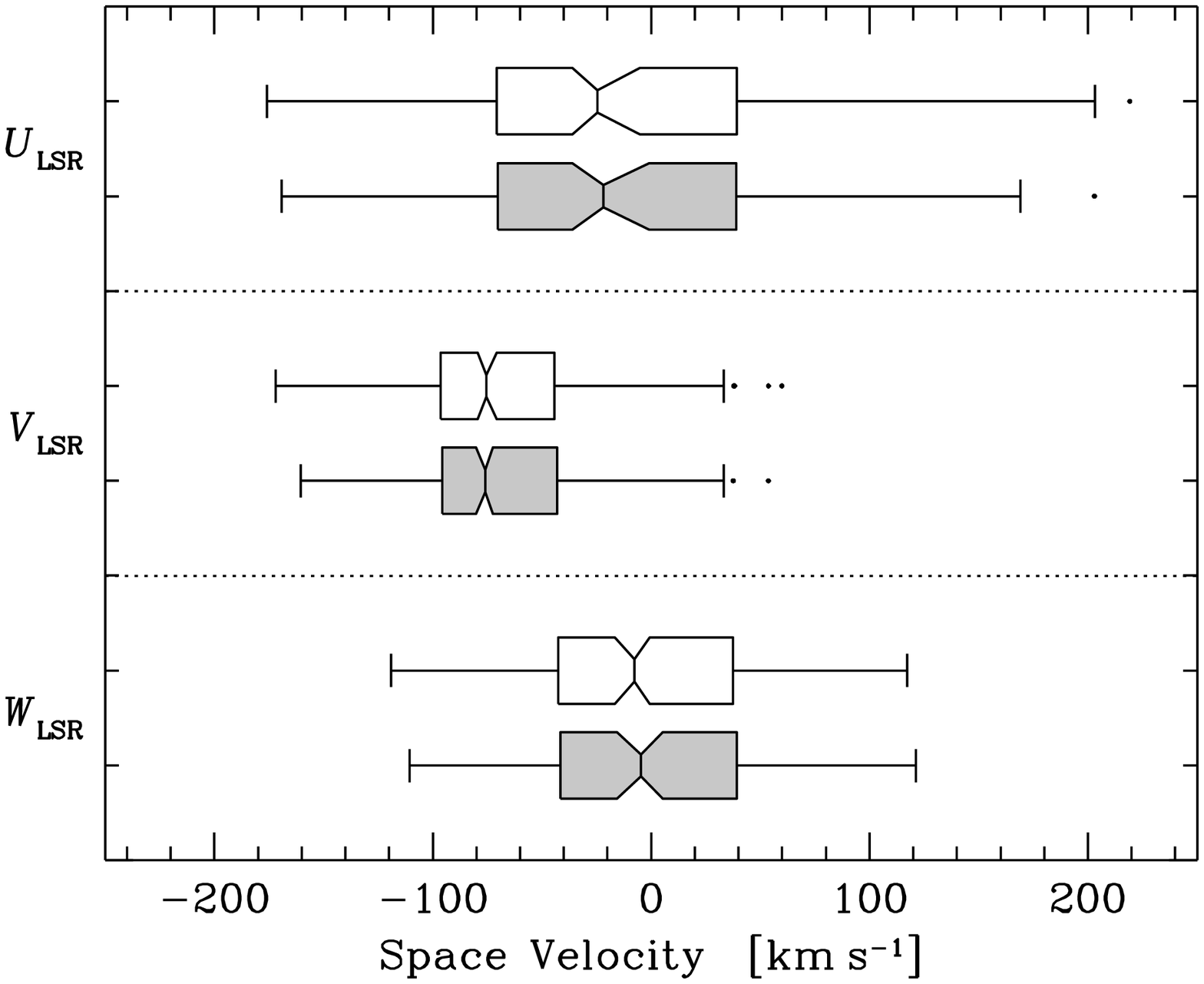}}
\caption{
        Boxplots showing the distribution in $(B-V)$, $M_{\rm V}$,
        $U_{\rm LSR}$, $V_{\rm LSR}$, and $W_{\rm LSR}$ for the
        full thick disk sample (295 stars, white boxes) and the thick disk
        sample with {\it uvby} photometry (229 stars, gray boxes).
        In the boxplots the central vertical line represents the
        median value. The lower and upper quartiles are represented by the
        outer edges of the boxes, i.e. the box encloses 50\,\% of the sample.
        The notches (waists) indicate the 95\,\% confidence intervals for the
        median value. The whiskers extend to the farthest data point that
        lies within 1.5 times the inter-quartile distance. Those
        stars that do not fall within the reach of the whiskers are regarded
        as outliers and are marked by solid circles.
	A common rule is that, samples for which the boxplot
        notch intervals do not overlap are likely to be different in their
        distributions.
        }
\label{fig:boxmvbv}
\end{figure}

A $TD/D$ ratio of 10 means that the star
is ten times more likely to be a thick disk star than a thin disk star.
For our previous studies of elemental abundances in the thick
and thin disks we selected stars with $TD/D>10$ as thick disk
stars and $TD/D<0.1$ as thin disk stars. In Bensby et al.~(\cite{bensby}) we
also analyzed spectra of a few stars with $TD/D$ between 10 and 2 and
it turns out that these stars show the typical thick disk trends
for the elemental abundances that the stars with $TD/D>10$ showed.
Thus it appears likely that we could use a somewhat lower $TD/D$ than
10 in order to increase our sample of
thick disk stars without compromising the results, but see discussion in
Sect.~\ref{sec:confusion}.

The most uncertain parameter in the calculation of the $TD/D$ ratios
is the normalization of the number density of thick disk stars in the
solar neighbourhood. 
Different authors quote different values:
2\,\% was found by Gilmore \& Reid~(\cite{gilmore}) and
Chen~(\cite{chen2}); $\sim6$\,\% was found by
Robin et al.~(\cite{robin}) and Buser et al.~(\cite{buser}); and
$\sim15$\,\%  by Chen et al.~(\cite{chen3}) and
Soubiran et al.~(\cite{soubiran}). 
As there is no simple way to determine which is the correct value we inspected
colour-magnitude diagrams (CMD) of the thin and thick disks, respectively,
derived using different normalizations for the solar neighbourhood thick
disk contribution (see Bensby et al.~submitted). From this inspection it is
clear that a very low normalization (2\,\%) must be ruled out since then the
thin disk CMD showed a population that was identical in age
and turn-off properties to that of the thick disk CMD.
The other values of the normalization are, however, not as easily
distinguished. As a compromise we will use a normalization of 10\,\% and
a $TD/D>2$ and $TD/H>1$ as our selection criteria for thick disk
stars.
The other parameters, such as the velocity dispersions, of course 
also influence the $TD/D$ ratios. They are however better known.

That the resulting samples are well defined regardless of the normalization
can be understood through the following argument.  
All stars with $TD/D\gtrsim2$, using a 10\,\% normalization,
will for all normalizations lower than 10\,\% still have $TD/D>1$.
This means that if we select stars with $TD/D\gtrsim2$ 
(10\,\% normalization) as our thick disk stars they would also be selected
as thick disk stars using any of the normalizations listed in the literature.
In the same way stars with $TD/D\lesssim0.6$ (10\,\% normalization)
would always be selected as thin disk irrespective of the value of the
thick disk normalization.

\begin{figure}
\centering
\resizebox{\hsize}{!}{\includegraphics[bb=18 144 592 482]{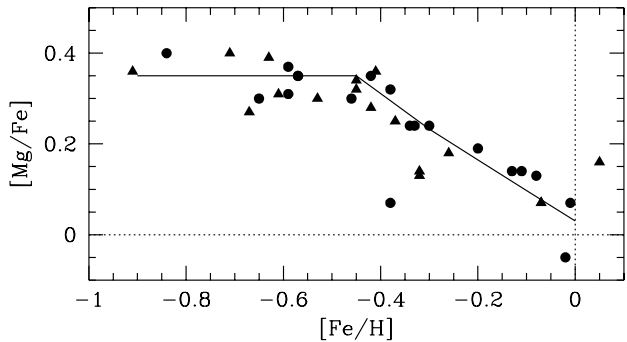}}
\caption{
        The [Mg/Fe] trend for a sample of thick disk stars from
        Bensby et al.~(\cite{bensby}) (circles) and
        Bensby et al.~(submitted) (triangles).
        The stars were selected according to
        the same kinematical criteria as discussed in Sect.~\ref{sect:kin}.
        The solid line indicates the degree of $\alpha$-enhancement that we
        have used at different [Fe/H]
        (see Table~\ref{tab:agemet}).
        }
\label{fig:alphaelem}
\end{figure}

\begin{figure}
\centering
\resizebox{\hsize}{!}{\includegraphics{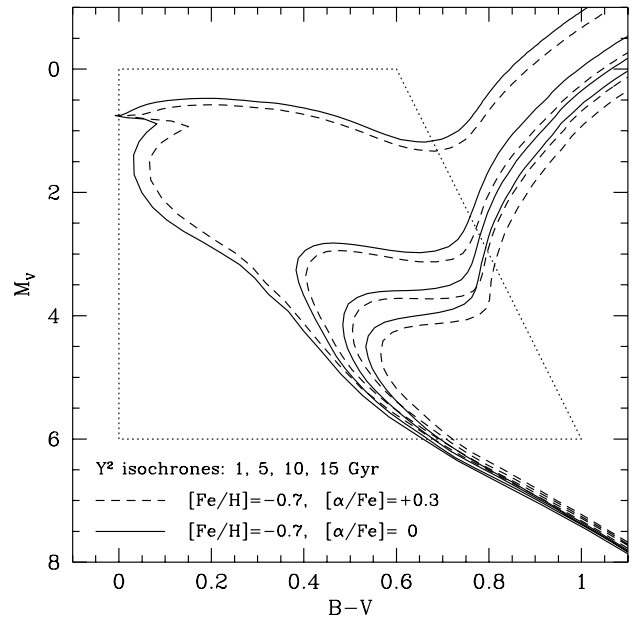}}
\caption{
        Example of how $\alpha$-enhancement changes the stellar
        isochrones. The plotted isochrones are the Y$^{2}$-isochrones from
        Kim et al.~(\cite{kim}) and Yi et al.~(\cite{yi2001}) and have 
	ages, [Fe/H], and $\alpha$-enhancements as indicated.
        The area enclosed by the dotted lines are the restrictions in 
	$(B-V)$ and $M_{\rm V}$
        discussed in Sect.~\ref{sect:kin}.
        }
\label{fig:isoalpha}
\end{figure}

We will restrict
the selection of the stellar sample to a certain area in 
the $(B-V)$-$M_{\rm V}$ plane.
The limits have been marked in 
Fig.~\ref{fig:isoalpha} by the dotted lines. 
This area has been selected on the grounds
that we do not want to include stars for which 
age determinations are nearly impossible,
which is the case the lower part of the main sequence 
where the isochrones are very crowded (i.e. $M_{\rm V}>6$). We also exclude
stars that lie on the giant branch, since the metallicity calibrations
are not valid for such stars (see Feltzing et al.~\cite{feltzing}).
In this region the colour of the stellar isochrones are also uncertain
(see e.g. Yi et al.~\cite{yi2001}, their Fig.~2).
We have also excluded stars that fall outside the following 
boundaries: $(B-V)<0$, $(B-V)>1$, and $M_{\rm V}<0$
(see Fig.~\ref{fig:isoalpha}).

Selecting all stars in the Hipparcos catalogue that have relative errors in
their parallaxes less than 25\,\% (and not flagged as binaries or probable 
binaries), that have radial velocities
published in the compilation by Barbier-Brossat et al.~(\cite{barbier}) we get
a sample of $\sim12\,600$ stars. Adding our thick disk criteria ($TD/D>2$
and $TD/H>1$) and the restrictions in $(B-V)$ and $M_{\rm V}$ given above,
we get a sample of 295 stars that are likely to belong to the Galactic thick 
disk.

\begin{figure*}
\centering
\resizebox{\hsize}{!}{
        \includegraphics[bb=18 144 565 718, clip]{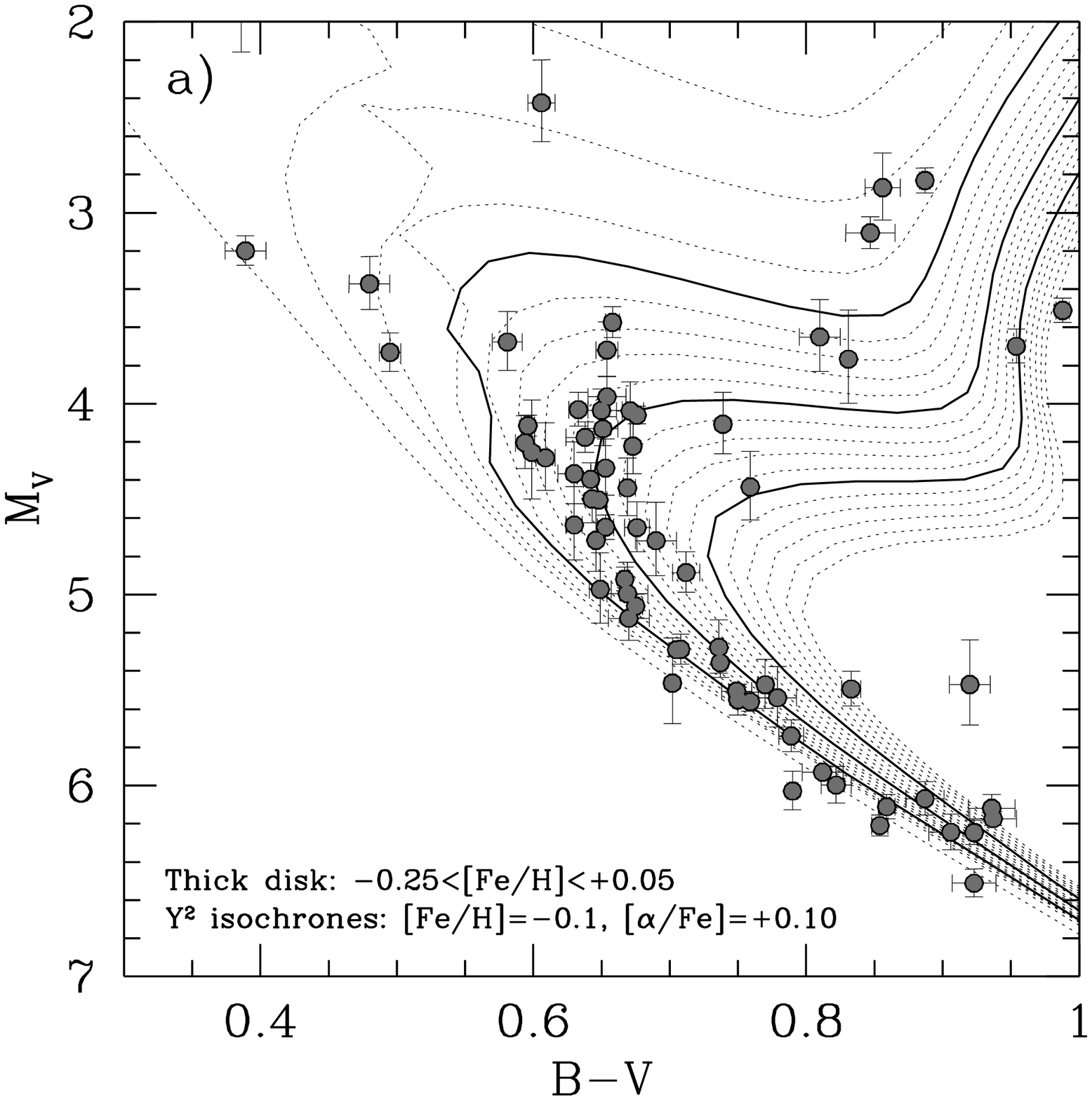}
        \includegraphics[bb=75 144 565 718, clip]{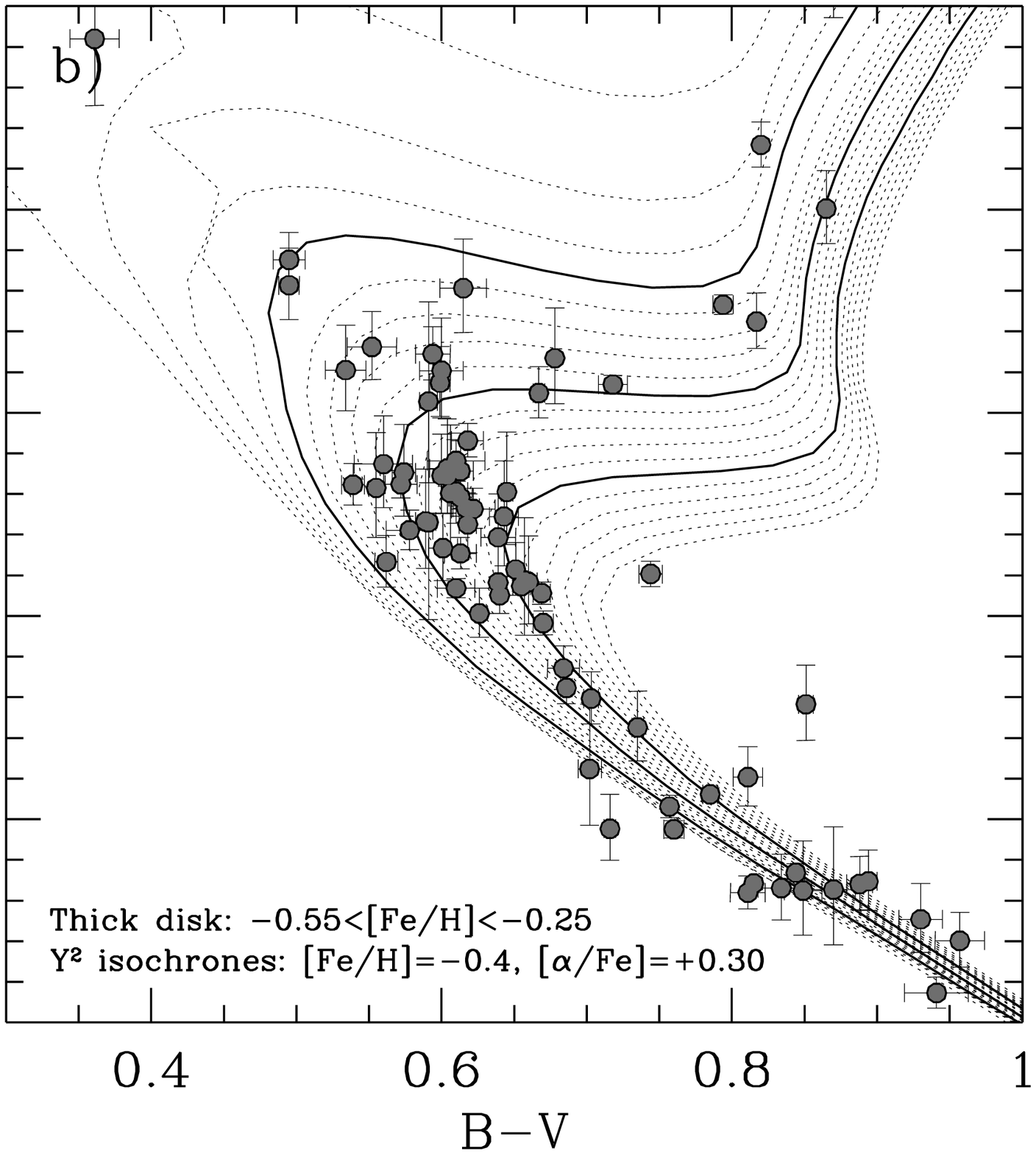}
        \includegraphics[bb=75 144 592 718, clip]{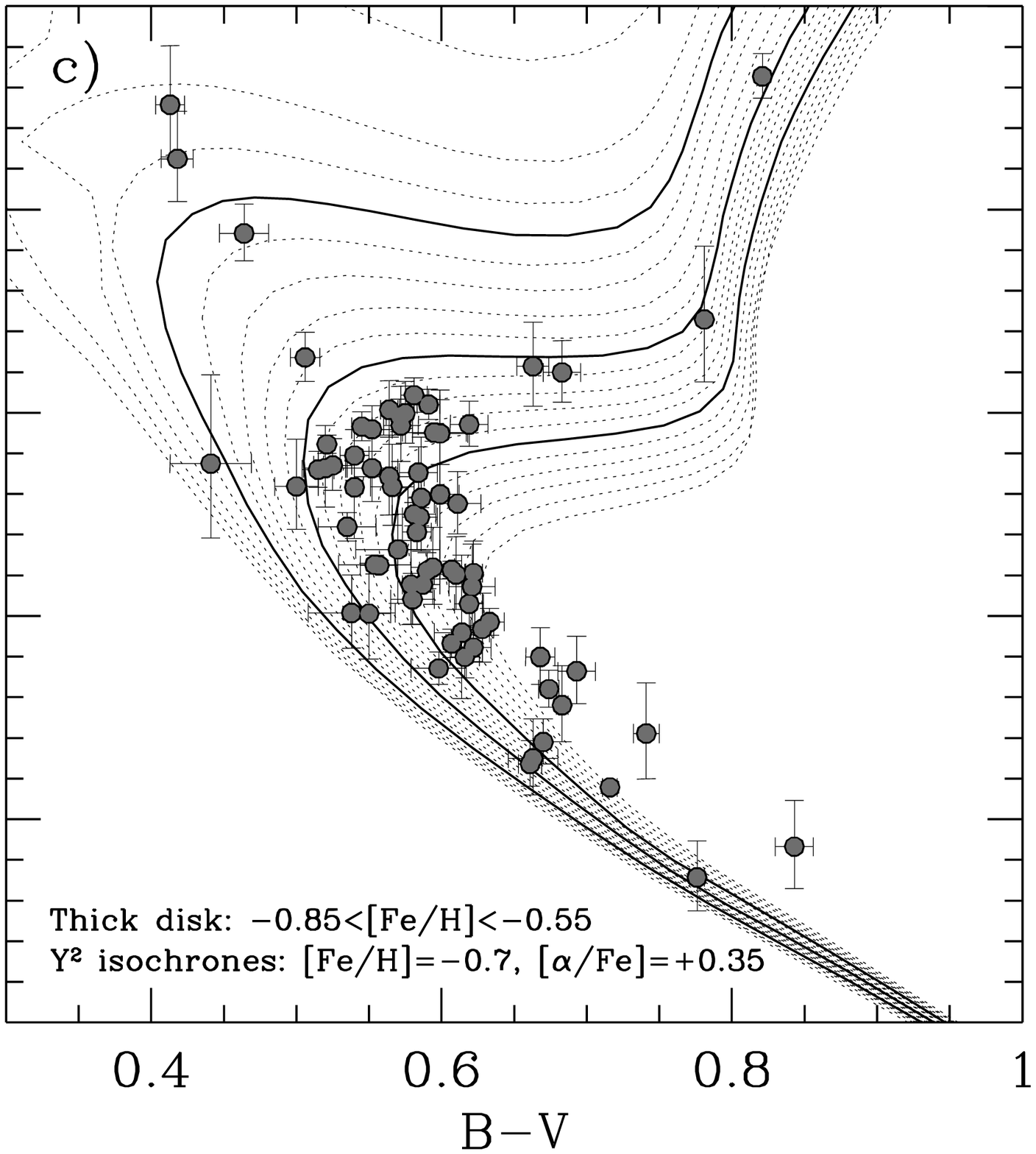}}
\caption{
        CMDs for the thick disk which show how the
        age distribution increases to older ages as we move to
        lower [Fe/H]. The isochrones are as indicated.
        We show ages from 20 to 1 Gyr with
        a step of 1 Gyr. The isochrones for 5, 10, 15, and 20 Gyrs are
        indicated by solid lines.
        }
\label{fig:hrthick}
\end{figure*}

Estimates of [Fe/H] for the stars can be derived
using calibrations of Str\"omgren photometry (compare Feltzing
et al.~\cite{feltzing}). However, not all the stars in the Hipparcos
catalogue have published Str\"omgren photometry.
229 out of the 295 thick disk stars have Str\"omgren photometry
from the compilation by Hauck \& Mermilliod~(\cite{hauck}).

We use the metallicities calculated for the catalogue used in
Feltzing \& Holmberg~(\cite{feltzing2000}) and
Feltzing et al.~(\cite{feltzing}). In Fig.~\ref{fig:boxmvbv}
we compare the distributions of $(B-V)$ and $M_{\rm V}$ and the distributions 
of the space velocities for the full thick disk sample (295 stars) and the 
thick disk sub-sample which has metallicities (229 stars).

At a quick glance, the distributions of the $M_{\rm V}$, $(B-V)$, and the 
$U_{\rm LSR}$, $V_{\rm LSR}$, $W_{\rm LSR}$ velocities do not seem to be 
appreciably affected when the stars without 
Str\"omgren photometry are rejected.  We verify this by performing two-sample
Kolmogorov-Smirnov (KS) tests. The probabilities that the distributions are 
identical is 99.4\,\% for $M_{\rm V}$, 89.7\,\% for $(B-V)$, and $>$\,99.9\,\% 
for all three velocities. This means that for all but $(B-V)$ the distributions
are essentially the same. For $(B-V)$ we have inspected the CMDs before
and after adding the Str\"omgren photometry. This inspection showed that
there is no appreciable deterioration in the sampling of red, old stars.
This makes us comfortable in believing that we do not introduce any
of the biases discussed in Feltzing et al.~(\cite{feltzing}).

In summary we do not introduce any bias to our thick disk sample by only 
selecting those stars with published Str\"omgren photometry.

\section{Abundances of $\alpha$-elements in the stars}
\label{sect:alpha}

Stars more metal-poor than the Sun have long been known to show enhanced levels
of $\alpha$-elements (see e.g. Edvardsson et al.~\cite{edvardsson}).
In Bensby et al.~(\cite{bensby}) and Bensby et al.~(submitted) we
use the same kinematic definitions of thick disk stars as we do here
and we derive stellar abundances for a large number of thin and thick disk
stars spanning a range of metallicities. In Fig.~\ref{fig:alphaelem} we show
the resulting [Mg/Fe] vs [Fe/H] trend for the thick disk stars in those
two papers. 

As can be seen from Fig.~\ref{fig:alphaelem} the enhancement of
[Mg/Fe] varies with [Fe/H] for the thick disk. At [Fe/H]\,$=$\,$-0.4$ 
the enhancement is 
$\sim0.3$\,dex while at [Fe/H]\,$=$\,0 it has decreased to almost solar
values. 
The thick disk stars in Mashonkina et al.~(\cite{mashonkina2003})
show the same degree of Mg enhancement, i.e. [Mg/Fe]\,$\approx 0.3$--0.4
for metallicities below [Fe/H]\,$=-0.4$.
(Note that the $\alpha$-element abundance often is defined
as the average of the Mg, Si, Ca, and Ti abundances, while
in Fig.~\ref{fig:alphaelem} we show Mg). 

It is important to know how much enhanced the stars are in
the $\alpha$-elements as $\alpha$-enhancement has a strong
effect on stellar evolutionary tracks, and hence on the stellar
isochrones that we want to use to determine the ages.
A set of isochrones with an $\alpha$-enhancement of [$\alpha$/Fe]\,$=$\,0.3
is compared to a set of solar-scaled isochrones in Fig.~\ref{fig:isoalpha}.
The effect of taking $\alpha$-enhancement into account is that ages
will be lower than otherwise.

In the determination of stellar ages from isochrones with different [Fe/H] 
we will use $\alpha$-enhancements that are in concordance with the the
trend that has been outlined in Fig.~\ref{fig:alphaelem} and is also
listed in Table~\ref{tab:agemet}. The Yonsei-Yale (Y$^{2}$) set of isochrones
(Yi et al.~\cite{yi2001}; Kim et al.~\cite{kim})
provides a versatile tool since they not only have published calculated
sets of isochrones but also an interpolator that enables the user to calculate 
a set of isochrones with a specific [Fe/H] and [$\alpha$/Fe].

\section{Age-metallicity relation in the thick disk}
\label{sec:amr}

Figure~\ref{fig:hrthick} shows three CMDs for stars with thick disk
kinematics and [Fe/H] derived from Str\"omgren photometry. Each CMD is 
centered at a different metallicity; [Fe/H]\,$=$\,$-0.7$, $-0.4$, and $-0.1$, 
respectively. Stellar isochrones with relevant metallicities and 
$\alpha$-enhancements (see Sect.~\ref{sect:alpha}) are also plotted. 

A visual inspection of these CMDs directly shows that the turn-off for the 
CMD centered at $-0.1$ dex is significantly younger than the turn-off in the 
CMD centered at $-0.7$ dex 
(see Figs.~\ref{fig:hrthick}a and \ref{fig:hrthick}c, respectively).

We now proceed to quantify this visual impression.
The stars are divided into 10 sub-samples according to their metallicities. 
Each sub-sample has a central metallicity and all stars $\pm0.15$\,dex 
around this value are included in the sub-sample. The central metallicity 
changes by 0.1\,dex between each sub-sample. In this way we create a sliding 
metallicity binning for the age determinations, i.e. the sub-samples are not 
independent. 

Ages for each sub-sample were simply estimated in the following way:
\begin{itemize}
\item 	A set of theoretical isochrones were generated
	according to the central metallicity and appropriate 
	$\alpha$-enhancement (see Sect.~\ref{sect:alpha}). 

\item 	The stars were plotted in the CMD
	together with the isochrones (see examples in 
	Fig.~\ref{fig:hrthick}).

\item 	The plot was inspected and an age was estimated
	for each star (within the restricted area marked out
	in Fig.~\ref{fig:isoalpha}).

\item 	The median age, lower, and upper quartiles where then
	calculated for each sub-sample.
\end{itemize}

\noindent
The results are collected in Table~\ref{tab:agemet}. The individual
stellar ages are good estimates of the stars age {\it given} the set of 
isochrones. More sophisticated methods to derive stellar ages
from isochrones exist (compare e.g. Ng \& Bertelli~\cite{ng};
Feltzing et al.~\cite{feltzing}; Pont \& Eyer~\cite{pont};
Rosenkilde J\o rgensen et al.~in prep.),
However, the age estimates we make are virtually identical to those
from the more sophisticated methods (e.g. Rosenkilde J\o rgensen who
uses a method akin to Pont \& Eyer~\cite{pont}). Errors on derived ages 
though are a different matter (Rosenkilde J\o rgensen private comm.).

\begin{table}
\caption{
        Median ages and spreads for the sub-samples.
        The first column gives the central
        metallicity, and the second column the $\alpha$-enhancement.
	Columns 3--6 give the follwoing for the samples
	selected with $TD/D>2$: the number of stars in the sub-sample,
        the lower quartile for the age distribution, the median age,
        and the upper quartile for the age distribution.
	Columns~7--10 give the same information but for the samples
	selected with $TD/D>10$.
        }
\label{tab:agemet}
\centering
\setlength{\tabcolsep}{1.5mm}
\begin{tabular}{rrrrrr|rrrr}
\hline\hline
\noalign{\smallskip}
[Fe/H]  & [$\alpha$/Fe] & $N$ &  \multicolumn{3}{c|}{Age (Gyr)} & $N$ & \multicolumn{3}{c}{Age (Gyr)} \\
        &               &     &  \multicolumn{3}{c|}{$TD/D>2$}  &     & \multicolumn{3}{c}{$TD/D>10$} \\
        &               &     & 1/4 &   1/2   & 3/4             &     & 1/4 &   1/2   & 3/4           \\
\noalign{\smallskip}
\hline
\noalign{\smallskip}
 $-$0.90 & +0.35 &   25  &     8.7  &   13.6  &  15.9           & 19  & 11.5  & 14.0  &  16.6         \\
 $-$0.80 & +0.35 &   39  &    11.5  &   13.5  &  16.2           & 31  & 12.5  & 13.5  &  16.0         \\
 $-$0.70 & +0.35 &   61  &    11.5  &   13.2  &  16.2           & 47  & 11.8  & 13.6  &  16.5         \\
 $-$0.60 & +0.35 &   65  &    11.1  &   13.0  &  15.0           & 44  & 11.2  & 13.0  &  14.9         \\
 $-$0.50 & +0.35 &   71  &     9.3  &   12.1  &  14.0           & 46  &  9.5  & 11.5  &  14.0         \\
 $-$0.40 & +0.30 &   58  &     8.7  &   11.7  &  13.5           & 39  &  8.5  & 11.2  &  13.8         \\
 $-$0.30 & +0.20 &   62  &     8.0  &   10.8  &  12.8           & 32  &  8.1  & 10.7  &  12.6         \\
 $-$0.20 & +0.15 &   60  &     7.4  &    9.5  &  10.5           & 26  &  7.8  &  9.6  &  10.6         \\
 $-$0.10 & +0.10 &   56  &     5.5  &    7.7  &   9.8           & 22  &  7.0  &  8.8  &  10.5         \\
    0.00 & +0.03 &   35  &     4.5  &    8.1  &  10.1           & 14  &  6.5  &  9.7  &  10.5         \\
\noalign{\smallskip}
\hline
\end{tabular}
\end{table}

\begin{figure}
\centering
\resizebox{\hsize}{!}{\includegraphics{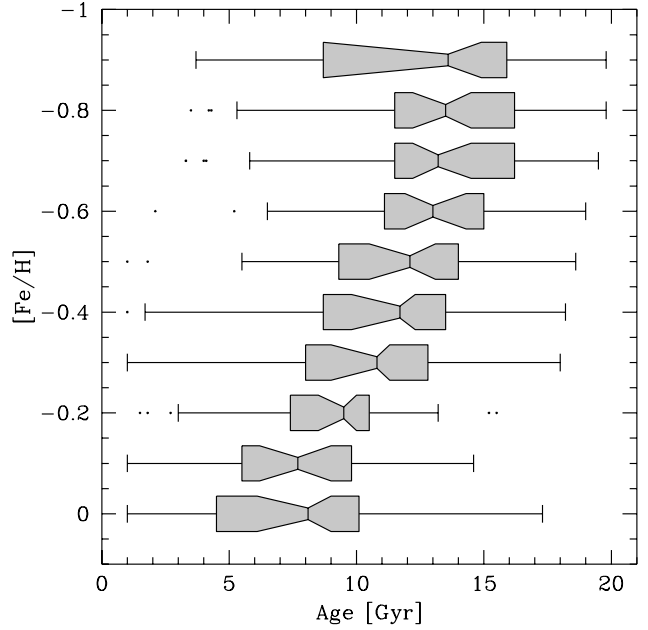}}
\caption{
        Age-metallicity diagram for the age estimates and central
        metallicities listed in Table~\ref{tab:agemet} for the samples
	selected with $TD/D>2$.
        We use boxplots to illustrate the age distributions within each
        sub-sample. For a description of box plots see caption of
        Fig.~\ref{fig:boxmvbv}.
        }
\label{fig:amr}
\end{figure}

The age estimates together with the 
central metallicities for each bin can now be plotted together in an 
age-metallicity diagram (see Fig.~\ref{fig:amr}). 
We see here clearly that indeed the central ages keeps decreasing as the 
metallicity increases. 

We believe that the {\it change} in age with [Fe/H] indeed is significant as we 
are using one set of isochrones and one sample of stars that have had their 
parameters determined in the same way. Thus we should not expect systematic
errors between bins. Furthermore the decline is up to 5 billion years
from the most metal-poor to the most metal-rich bin. Such a large, and 
systematic, change would be hard to achieve through an erroneous analysis.

\section{Thin disk -- thick disk confusion?}
\label{sec:confusion}

\begin{figure}
\centering
\resizebox{\hsize}{!}{\includegraphics{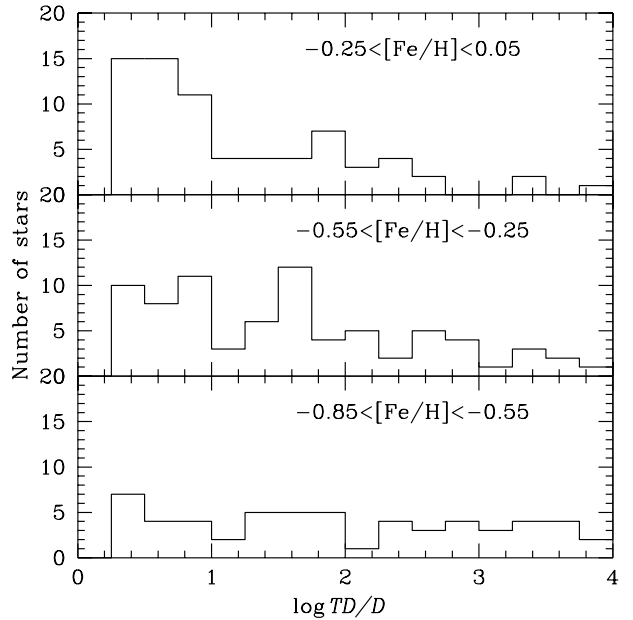}}
\caption{
	Histograms showing the distributions of the $TD/D$ ratios
	in three different metallicity ranges for all
	stars having $TD/D>2$ (same stars as in Fig.~\ref{fig:hrthick}).
        }
\label{fig:tdd_distribution}
\end{figure}

In Fig.~\ref{fig:amr} we see  a strong decrease in the mean age for 
the highest metallicities. 
Given that the thin disk dominates more and more as we go to 
higher metallicities the question arises: are we simply adding more
and more of the younger thin disk stars into our sample?

There is in fact some hints in our data that we are picking up thin
disk stars in our most metal-rich bins. Figure~\ref{fig:tdd_distribution}
shows the distributions of $TD/D$ for the three CMDs in Fig.~\ref{fig:hrthick}.
As can be seen we are picking up more and more stars
with low $TD/D$ ratios as we go to higher metallicities.           
The stars with $TD/D<10$ could be intervening thin disk stars.

Ideally, if the metallicity distribution functions for the two disks
were well known we could weight the probabilities we use to select the stars
accordingly.

The metallicity distributions of 
the thin and thick disks peak at different metallicities 
(Wyse \& Gilmore~\cite{wyse}). 
The thick disk appears to peak in the interval $\rm -0.7 < [Fe/H] < -0.5$
and the thin disk around $-0.2$\,dex.
However, the two distributions  are not well constrained. Especially the
distribution for the thick disk truncates at $\rm [Fe/H]\approx -0.2$ 
in Wyse \& Gilmore~(\cite{wyse}). We find that due to the shortcomings
in our knowledge about the metallicity distribution functions
of the two disk populations it is not possible to use them to further
deconvolve the two populations.

\begin{figure*}
\centering
\resizebox{\hsize}{!}{
        \includegraphics[bb=18 144 565 718, clip]{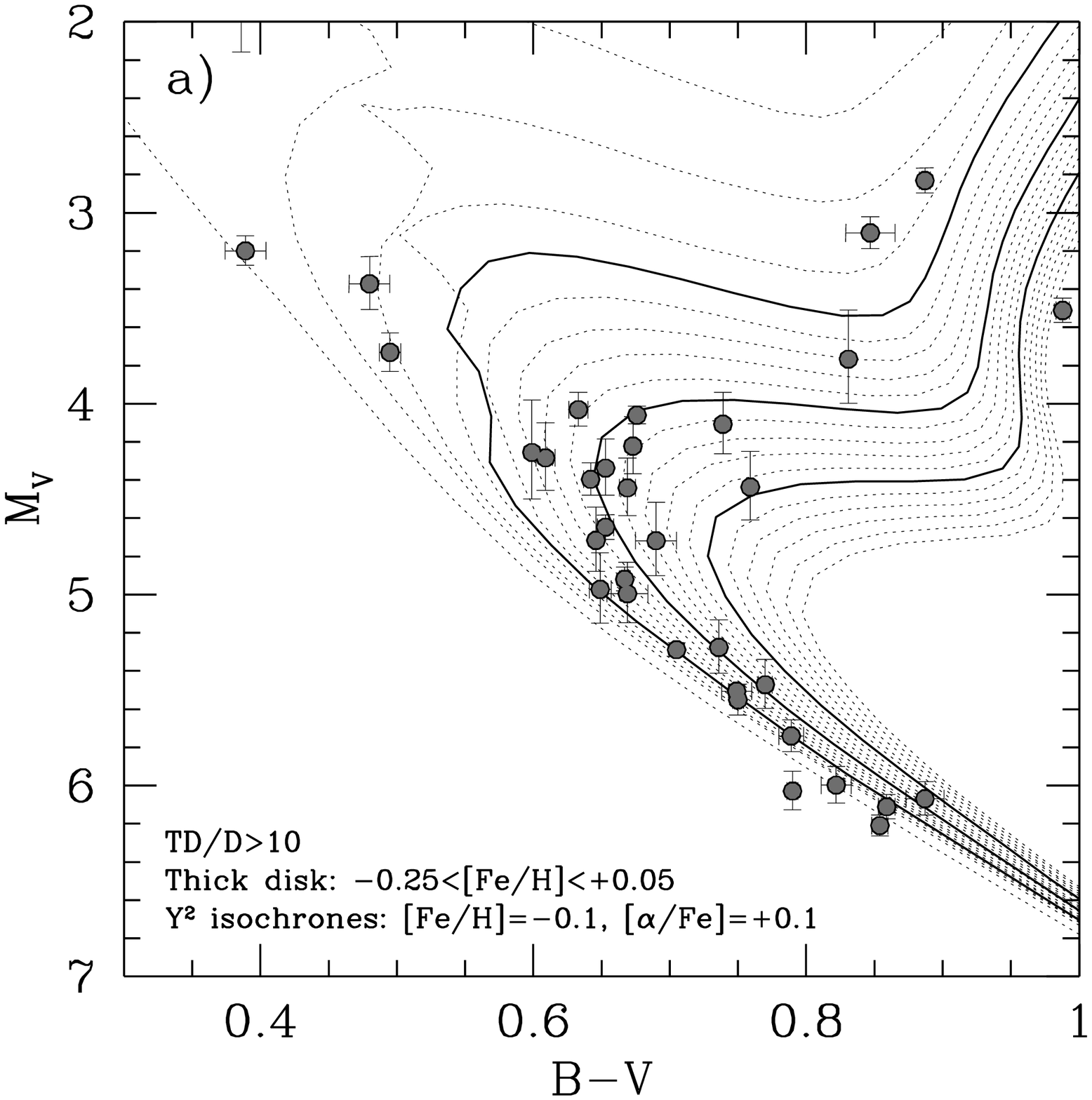}
        \includegraphics[bb=75 144 565 718, clip]{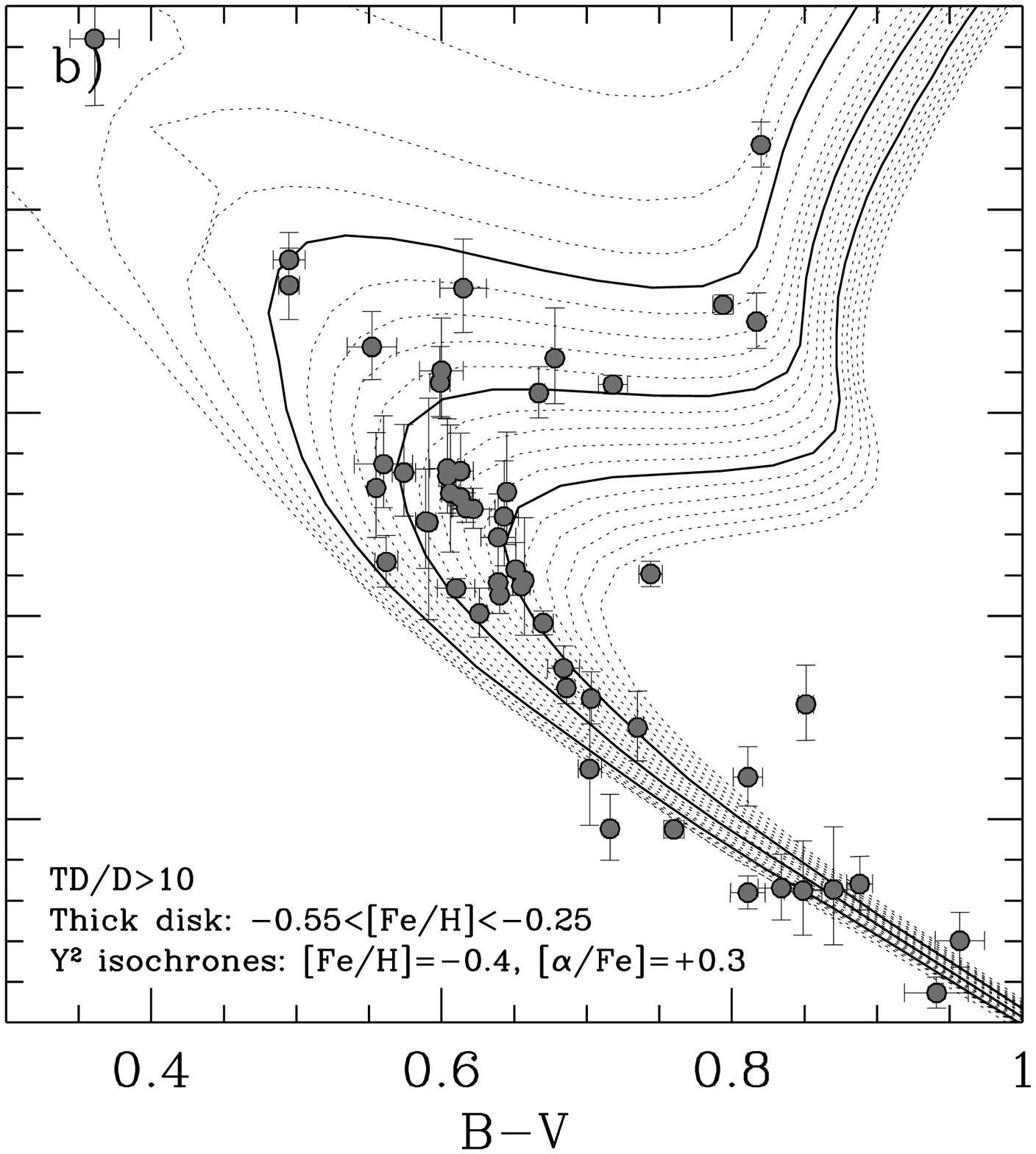}
        \includegraphics[bb=75 144 592 718, clip]{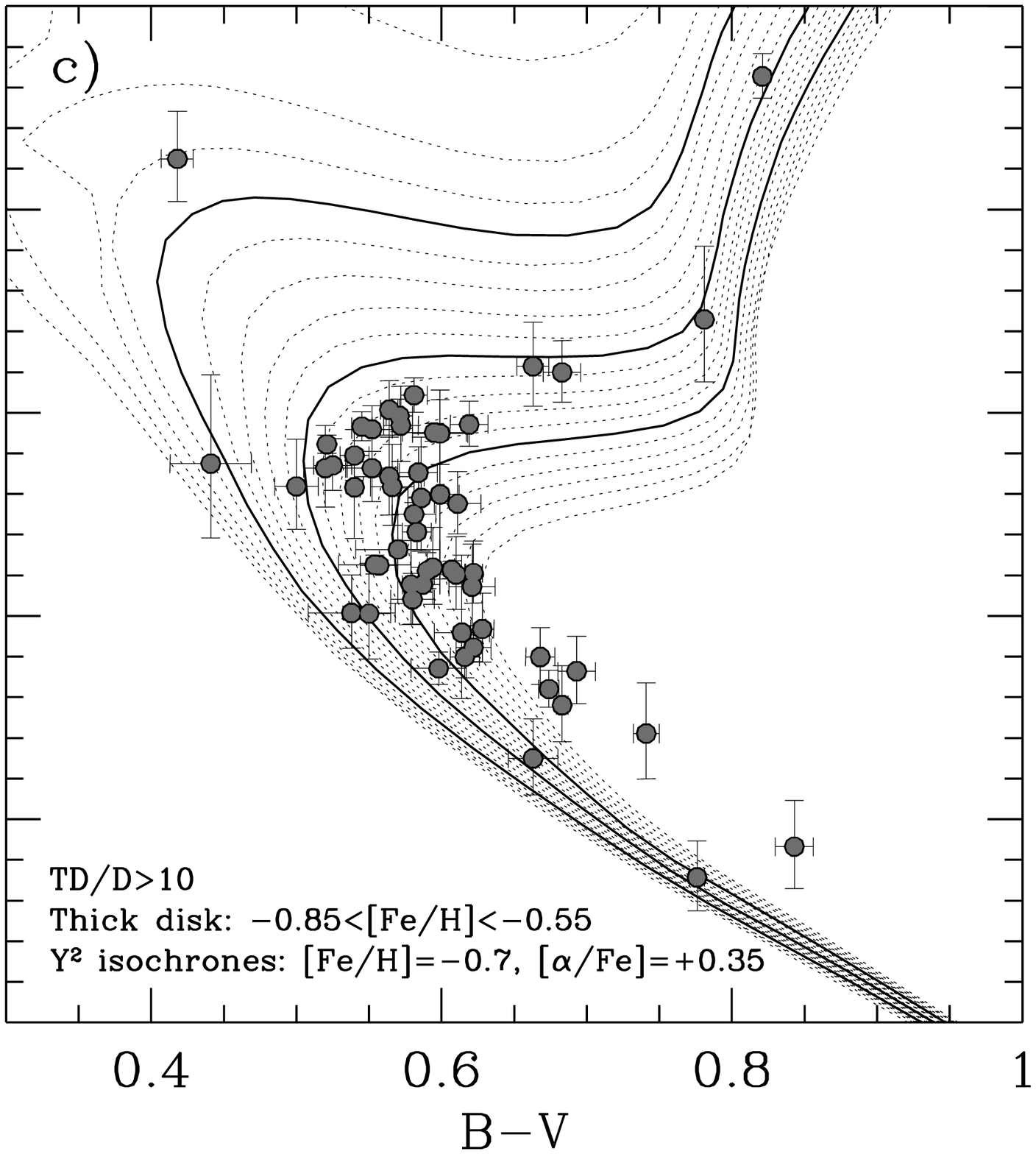}}
\caption{
	Same as Fig.~\ref{fig:hrthick} but using $TD/D>10$ instead.
        }
\label{fig:hrthick_tdd10}
\end{figure*}

\begin{figure}
\centering
\resizebox{\hsize}{!}{\includegraphics{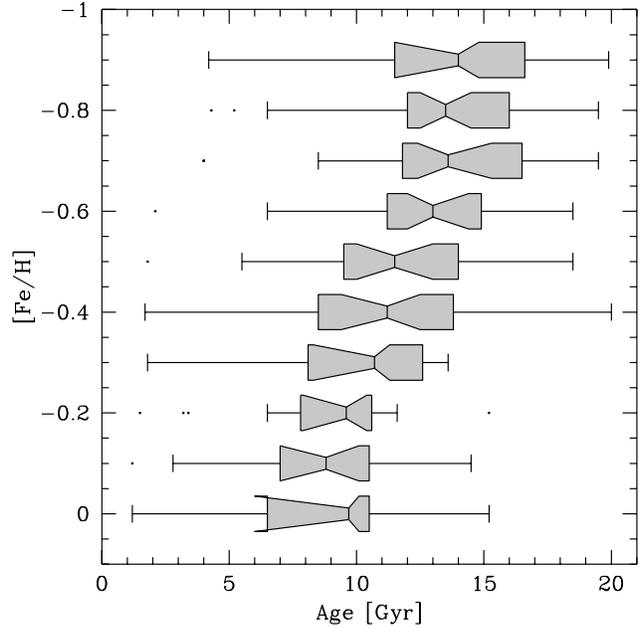}}
\caption{
	Same as Fig.~\ref{fig:amr} but using $TD/D>10$ instead.
        }
\label{fig:amr_tdd10}
\end{figure}

We will here try to address the issue of intervening thin disk stars in a 
different way. In Fig.~\ref{fig:hrthick_tdd10} we show the same 
CMDs as in Fig.~\ref{fig:hrthick} but with a stronger constraint
on $TD/D$, i.e. $TD/D>10$, and compare Fig~\ref{fig:tdd_distribution}.
In this figure we see the same trend (albeit with fewer stars) 
as in the original CMDs, strengthening
our earlier conclusion that there is an AMR present in the thick disk.
In Fig.~\ref{fig:amr_tdd10} we show the resulting AMR constructed in the 
same way as in Fig.~\ref{fig:amr}. Note, however, in this new case we have
as a furher precaution excluded all stars with $M_{\rm V}>5.4$ in order
to make sure that we have as good ages as possible. Furthermore, the reader
should note that very few stars are on the sub-giant branch, where
the evolutionary timescale is short and hence age determinations can
be erroneous (see Pont \& Eyer~\cite{pont}). As a further extreme test 
we have also
inspected the CMDs where also all star with $V_{\rm LSR}>-50$\,km\,s$^{-1}$
have been excluded. These CMDs clearly shows that the stars centered
on [Fe/H]\,$=-0.4$ form in the mean a younger population that the
stars centered around [Fe/H]\,$=-0.7$.

Thus we again find an AMR to be present in the thick disk.

\section{Discussion}
\label{sec:discussion}

\begin{figure}
\centering
\resizebox{\hsize}{!}{
        \includegraphics[bb=18 144 592 475,clip]{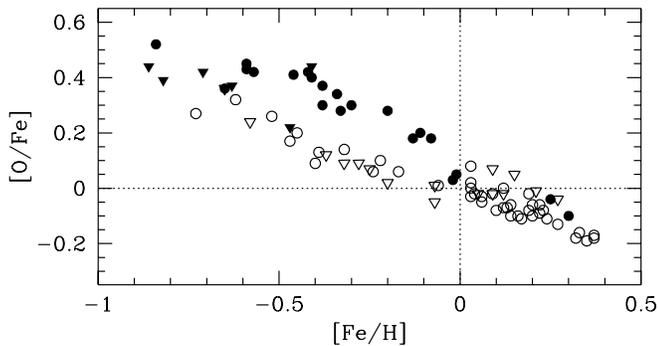}}
\caption{
        Abundance trends for oxygen in the thin and thick disks
        (Bensby et al.~\cite{bensby_syre}). Thin and thick disk
        stars are marked by open and filled symbols, respectively.
        Triangles are data points that have been gathered from
        Nissen et al.~(\cite{nissen}).
        }
\label{fig:syre}
\end{figure}

From a spectroscopist's point of view there is now mounting evidence that the
stars that today have kinematics that we associate with the thick disk show
elemental abundance trends that are distinct from those seen in stars on solar
type orbits (thin disk). A recent review can be found in 
Nissen~(\cite{nissen_carnegie}). Our own results have been published in 
Feltzing et al~(\cite{feltzing2003}, \cite{feltzing_carnegie}) and
Bensby et al.~(\cite{bensby}, \cite{bensby_syre}), 
see also Fig.~\ref{fig:syre}. 
These observations appear to point to some sort of homogeneity in the chemical 
enrichment process that proceeded the formation of the stars that we today 
associate with the thick disk.

Our new result that there is a possible age-metallicity relation in the thick 
disk, adds a further dimension. The results indicate that star formation 
continued up to 5 billion years in the population we identify 
as the thick disk.

The observational properties and a standard interpretation for the chemical 
evolution of the thick disk can be described in the following way; 
The abundance trends for the thick disk are well-defined and show large 
$\alpha$-enhancements at a constant value for metallicities below 
[Fe/H]\,$\approx$\,$-0.4$. This is normally interpreted as that the star 
formation was intense and that massive stars were the main contributors to the 
chemical enrichment. At higher [Fe/H] the $\alpha$-enhancement starts to 
decline toward solar values. This is typical for what happens when long-lived 
low-mass stars starts to contribute to the chemical enrichment through the 
explosion of SN\,Ia. The SN\,Ia mainly produce iron peak elements and none 
or only little of the $\alpha$-elements which results in a lowering of the 
[$\alpha$/Fe] ratio (compare e.g. Fig.~8.6 in Pagel~\cite{pagel}).

The time at which the decline starts is not only a function of the lifetime 
of SN\,Ia but also of the SN\,Ia rate. The exact model, and hence lifetime 
for SN\,Ia, is still debated. The two most probable scenarios are a double 
or a single degenerate system consisting of either two white dwarfs or a 
white dwarf and a red giant. In the latter case the life time is set by the  
main sequence lifetime of the star that becomes the red giant, while in the 
first case the lifetime for the system prior to become supernova is set by the 
time it takes the two white dwarfs to spiral in and coalesce, 
and hence it depends on the 
initial separation and could be longer than the Hubble time. For a recent,
in-depth discussion of these issues see Livio~(\cite{livio}). Furthermore, 
the population synthesis of the SN\,Ia progenitors do, in general, not agree 
too well with the observed frequencies of possible progenitor systems 
(see Livio~\cite{livio} and references therein). This means that any
interpretation of our $\alpha$-abundance trends in terms of a minimum
timescale for the star formation period in the thick disk will be
rather complex (see also Fig. 5.7 in Matteucci~\cite{matteucci}).

However, our investigation of the relation between ages, metallicities,
and elemental abundances for stars with typical thick disk kinematics points 
to a conclusion that the SN\,Ia {\it rate} peaked after a few 
($\sim$3--4) billion years.
We also note that the observed abundance trends are tight,
compare the oxygen trends in Fig.~\ref{fig:syre}, which should 
indicate that the gas must have been rather well mixed throughout.
This seems to indicate that the gas that the thick disk stars formed out
of must have been confined to a reasonably ``small" physical volume in order
for the mixing to work efficiently.

Simulations of galaxy formation indicate that we should expect all or a large 
part of our galaxy to have been assembled from smaller parts 
(e.g. Murali et al.~\cite{murali}). The question then arises -- did these 
smaller building blocks form stars before they merged into today's thick disk? 
If they did and if they were of different sizes, as one may imagine, then they 
would each form their own unique set of elemental abundance trends which would  
depend on the initial mass function and the star formation rate in each 
individual building block. Then clearly we would not see any homogeneous 
elemental abundance trends as has been found by several authors
(Fuhrmann~\cite{fuhrmann}; Mashonkina \& Gehren~\cite{mashonkina};
Prochaska et al.~\cite{prochaska}; Feltzing et al.~\cite{feltzing2003}).  
Of course, if these building blocks were collected before they formed stars it 
might be possible to find abundance trends in today's thick disk. 
Another, and perhaps 
more likely possibility, is that what we today observe as a thick disk 
originally was a thin disk that, after having produced its stars, was puffed 
up (see e.g. Robin et al.~\cite{robin}; Quinn et al.~\cite{quinn}). 
Such a scenario would allow the gas to be more well 
mixed at the time of star formation and abundance trends would be possible. 

Another possible scenario is a close encounter between the 
Milky Way and another galaxy. This scenario has been
modelled by Kroupa~(\cite{kroupa}) who shows that such an event 
would result in a kinematical heating of a pre-existing gaseous disk and
an increased star formation in this gas.

\section{Summary}
\label{sec:summary}

Using a sample of 229 kinematically selected thick disk stars
we have been able to probe the existence of an age-metallicity relation in
Galactic thick disk. Ages have been determined in a consistent way,
using isochrones with the appropriate levels of $\alpha$-enhancement at 
different metallicities as indicated by recent studies that 
use detailed abundance analysis. 
Although the stellar sample is rather small we believe
it to be free from severe biases that could affect the results. From
the investigation we are able to draw three main conclusions:
\begin{enumerate}
\item 	There is an age-metallicity relation present in the Galactic
	thick disk, indicating that it has had an ongoing star formation 
	for a time-period of up to 5\,Gyr (this is model dependent).
\item	The thick disk age-metallicity relation in combination with the 
	abundance trends for $\alpha$-elements in the thick disk, that show 
	signatures from SN\,Ia, indicate that the time-scale for the peak of
	the SN\,Ia rate in the thick disk is of the order 3--4\,Gyr.
\item	The quite long star formation period in the thick disk 
	strengthens the hypothesis that the thick disk formed as a result
	of an ancient merger event between the Milky Way and a companion 
	galaxy.
\end{enumerate}

Studies of the thick disk using nearby stars will always be be subject
to uncertainties due to the overlapping velocity-, and metallicity 
distributions of the thin and thick disk. At distances well above
$Z\approx1.5$\,kpc from the Galactic plane the thick disk is the dominant
stellar population. Determining accurate metallicities and 
$\alpha$-abundances for a larger sample of dwarf stars at high 
$Z$ would therefore enable accurate age determinations
that would verify the existence (or non-existence) of the thick disk AMR
deduced from nearby stars. 

\acknowledgements{
We would like to thank Poul Erik Nissen and the anonymous referee
for valuable comments on the first version of the paper.



\begin{thebibliography}{}

\bibitem[1994]{barbier}
 Barbier-Brossat, M., Petit, M., \& Figon, P., 1994, A\&AS, 108, 603

\bibitem[2003]{bensby}  
 Bensby, T., Feltzing, S., \& Lundstr\"om, I, 2003, A\&A, 410, 527

\bibitem[2004]{bensby_syre} 
 Bensby, T., Feltzing, S., \& Lundstr\"om, I, 2004, A\&A, 415, 155

\bibitem[1999]{berczik}  
 Berczik, P., 1999, A\&A, 348, 371

\bibitem[1999]{buser}
 Buser, R., Rong, J., \& Karaali, S., 1999, A\&A, 348, 98

\bibitem[1997]{chen2}
 Chen, B., 1997, ApJ, 491, 181

\bibitem[2001]{chen3} 
 Chen, B., Stoughton, C., Smith, J.A., et al., 2001, ApJ, 553, 184

\bibitem[2003]{chiappini}
 Chiappini, C., Romano, D., \& Matteucci, F., 2003, MNRAS, 339, 63

\bibitem[1993]{edvardsson} 
 Edvardsson, B., Andersen, J., Gustafsson, B., Lambert, D.L.,
 Nissen, P.E., \& Tomkin, J., 1993, A\&A, 275, 101

\bibitem[1998]{feltzing1998}
 Feltzing, S., \& Gustafsson, B., 1998, A\&AS, 129, 237

\bibitem[2000]{feltzing2000} 
 Feltzing, S., \& Holmberg, J., 2000, A\&A, 357, 153

\bibitem[2001]{feltzing} 
 Feltzing, S., Holmberg, J., \& Hurley, J.R., 2001, A\&A, 377, 911

\bibitem[2003a]{feltzing2003}
 Feltzing, S., Bensby, T., \& Lundstr\"om, I., 2003a, A\&A, 397, L1

\bibitem[2003b]{feltzing_carnegie}
 Feltzing, S., Bensby, T., Gesse, S., \& Lundstr\"om, I., 2003b
 Carnegie Observatories Astrophysics Series, Vol. 4:
 Origin and Evolution of the Elements, ed. A. McWilliam and M. Rauch
 (Pasadena: Carnegie Observatories,
  \texttt{http://www.ociw.edu/ociw/symposia/series/symposium4\\
        /proceedings.html})

\bibitem[1998]{fuhrmann}
 Fuhrmann, K., 1998, A\&A, 338, 161

\bibitem[1983]{gilmore}
 Gilmore, G., \& Reid, N., 1983, MNRAS, 202, 1025

\bibitem[1989]{gilmore2}
 Gilmore, G., Wyse, R.F.G., \& Kuijken, K., 1989, ARAA, 27, 555

\bibitem[1998]{hauck}
 Hauck, B., \& Mermilliod, M., 1998, A\&AS, 129, 431

\bibitem[2002]{kim}  
 Kim, Y.-C., Demarque, P., Yi, S.K., \& Alexander, D.R., 2002, ApJS, 143, 499 

\bibitem[2002]{kroupa}
 Kroupa, P., 2002, MNRAS, 330, 707

\bibitem[2001]{livio} 
 Livio, M., 2001, Supernovae and gamma-ray bursts: the greatest explosions 
 since the Big Bang, (eds) M. Livio, N. Panagia, K. Sahu.,
 STScI symp. ser., Vol. 13, p. 334, Cambridge University Press
 (astro-ph/0005344)

\bibitem[2001]{mashonkina}
 Mashonkina, L., \& Gehren, T., 2001, A\&A, 376, 232

\bibitem[2003]{mashonkina2003}
 Mashonkina, L., Gehren, T., Travaglio, C., \& Borkova, T., 2003, 
 A\&A, 397, 275

\bibitem[2001]{matteucci} 
 Matteucci, F., 2001, The Chemical Evolution of the Galaxy, 
 Astrophysics and Space Science Library, Volume 253, Dordrecht,
 Kluwer Academic Publishers

\bibitem[1991]{meusinger} 
 Meusinger, H., Stecklum, B., \& Reimann, H.-G., 1991, A\&A, 245, 57

\bibitem[2002]{murali}
 Murali, C., Katz, N., Hernquist, L., Weinberg, D.H., \& Dav\'e, R.,
 2002, ApJ, 571, 1

\bibitem[1998]{ng}
 Ng, Y.K., \& Bertelli, G., 1998, A\&A, 329, 943

\bibitem[2002]{nissen}
 Nissen, P.E., Primas, F., Asplund, M., \& Lambert, D.L.,
 2002, A\&A, 390, 235

\bibitem[2003]{nissen_carnegie}
 Nissen, P.E., 2003,
 Carnegie Observatories Astrophysics Series, Vol. 4:
 Origin and Evolution of the Elements, ed. A. McWilliam and M. Rauch
 (Pasadena: Carnegie Observatories,
  \texttt{http://www.ociw.edu/ociw/symposia/series/symposium4\\
        /proceedings.html})

\bibitem[1997]{pagel}
 Pagel, B.E.J., 1997, Nucleosynthesis and chemical evolution
 of galaxies, Cambridge University Press

\bibitem[1996]{pilyugin} 
 Pilyugin, L.S., \& Edmunds, M.G., 1996, A\&A, 313, 792

\bibitem[2004]{pont}
 Pont, F., \& Eyer, L., 2004, MNRAS, in press, (astro-ph/0401418)

\bibitem[2000]{prochaska}
 Prochaska, J.X., Naumov, S.O., Carney, B.W., McWilliam, A., \& Wolfe, A.M.,
 2000, ApJ, 120, 2513

\bibitem[1993]{quinn} 
 Quinn, P.J., Hernquist, L., \& Fullagar, D.P., 1993, ApJ, 403, 74

\bibitem[1996]{raiteri} 
 Raiteri, C.M., Villata, M., \& Navarro, J.F., 1996, A\&A, 315, 105

\bibitem[2003]{reddy}
 Reddy, B.E., Tomkin, J., Lambert, D.L., \& Allende Prieto, C., 2003, MNRAS,
 340, 304

\bibitem[2001]{reyle}
 Reyl\'e, C., \& Robin, A.C., 2001, A\&A, 373, 886

\bibitem[1996]{robin}
 Robin, A.C., Haywood, M., Cr\'eze, M., Ojha, D.K., \& Bienaym\'e, O., 1996,
 A\&A, 305, 125

\bibitem[2000]{rochapinto} 
 Rocha-Pinto, H.J., Maciel, W.J., Scalo, J., \& Flynn, C., 2000, A\&A, 358, 850

\bibitem[2003]{soubiran}
 Soubiran, C., Bienaym\'e, O., \& Siebert, A., 2003, A\&A, 398, 141

\bibitem[1980a]{twaroga} 
 Twarog, B.A., 1980a, ApJS, 44, 1

\bibitem[1980b]{twarogb} 
 Twarog, B.A., 1980b, ApJ, 242, 242

\bibitem[2001]{yi2001} 
 Yi, S., Demarque, P., Kim, Y.-C., Lee, Y.-W., Ree, C.H., 
 Lejeune, T., \& Barnes, S., 2001, ApJS, 136, 417 


\bibitem[1995]{wyse} 
 Wyse, R.F.G., \& Gilmore, G., 1995, AJ, 110, 2771




\end{thebibliography}
\end{document}